\documentclass{article}
\usepackage[utf8]{inputenc}
\usepackage{natbib}
\usepackage{soul}
\usepackage{xcolor}
\usepackage{hyperref}
\usepackage{setspace}

\title{Is the Reality Criterion Analytic?}
\author{David Glick\footnote{Stellenbosch Institute for Advanced Study (STIAS), Wallenberg Research Centre at Stellenbosch University, Stellenbosch 7600, South Africa} \footnote{Department of Philosophy, University of Sydney, NSW 2006, Australia} \and Florian J.\ Boge\footnotemark[1] \footnote{Institute for Theoretical Particle Physics and Cosmology, RWTH Aachen University, 52074 Aachen, Germany}}
\date{}

\begin{document}

\maketitle

\begin{abstract}
    Tim Maudlin has claimed that EPR's Reality Criterion is analytically true. We argue that it is not. Moreover, one may be a subjectivist about quantum probabilities without giving up on objective physical reality. Thus, would-be detractors must reject QBism and other epistemic approaches to quantum theory on other grounds.
\end{abstract}

\doublespace
\section{Introduction}
The Reality Criterion plays a central role in Einstein, Podolsky, and Rosen's argument against the completeness of quantum mechanics (QM).\footnote{Whether the Reality Criterion is needed to make EPR's point is another question. In his own (individual) presentations of the argument, Einstein does not include the Reality Criterion, which may cast doubt on his commitment to the principle \citep[p.62]{FineShakyGame}.}

\begin{description}
\item[Reality Criterion (RC):] If, without in any way disturbing a system, we can predict with certainty (i.e., with probability equal to unity) the value of a physical quantity, then there exists an element of reality corresponding to that quantity. \citep[p.777]{EPR}
\end{description}

RC shares much of the intuitive appeal of scientific realism. On the basis of certain predictions, it licenses claims about objective reality, which may then be used to explain our successful predictive practices. But, unlike scientific realism, RC only appeals to specific kinds of predictions made with \emph{certainty}. It then allows us to infer the existence of something in reality \emph{corresponding} to the predicted value of the quantity in question, e.g., some property of the system under study. 

By focusing on predictions that can be made with certainty, which EPR identify with probability one assignments, RC represents a substantially weaker connection between our predictions and the world than full-blown scientific realism. RC requires ontological commitment to substantially less than the full content of quantum theory that traditional realism would seem to endorse. Yet, certain proposed interpretations of QM must reject RC. QBism (which originally stood for ``quantum Bayesianism''), for instance, maintains that the probabilities delivered by QM correspond to agents' subjective degrees of belief about their future experiences (e.g., of measurement outcomes). Hence, there is the possibility, on QBism, to be perfectly certain about the outcome of some measurement---even assign it a probability equal to unity---without there being anything in reality that corresponds to the predicted outcome \citep[cf.,][]{FuchsMerminSchackIntroQBism}. 

Maudlin \citeyearpar{MaudlinBell}, however, claims that RC is not merely reasonable (as EPR find it), but analytically true:

\begin{quote}
...[RC] is, in the parlance of philosophers, \emph{analytic}. That is, this criterion follows just from the very meanings of the words used in it. The difference is this:\ one \emph{can} coherently (but not reasonably!) deny a merely reasonable claim, but one can't coherently deny an analytic proposition. \citep[p.6, original emphasis]{MaudlinBell}
\end{quote}

Analytic claims are true or false in virtue of meanings alone. To give the standard example, one can see that all bachelors are unmarried without any empirical knowledge---its truth simply follows from what we mean by ``bachelor.'' If RC is analytically true, as Maudlin maintains, its denial is incoherent. To reject RC is to fail to grasp the meanings of the terms involved.

\citet[p.\ 6]{WernerComment} rightly points out that RC is a \emph{criterion}, i.e., a sufficient condition for identifying elements of reality, and that analyticity is something that traditionally applies to statements or judgments. However, consider the following conditional:\ ``for all men, if we can ascertain that they are unmarried, then we may infer that they are bachelors.'' This conditional  provides a criterion for checking whether (grown) men are indeed bachelors:\ just investigate their marital status. But like the corresponding statement, the criterion itself might be called ``analytic,'' as denying its applicability would imply a want of understanding of the terms involved.

%The main aim of \citet{MaudlinBell} is to argue that, in light of Bell's theorem, physical reality is non-local. His claim that RC is analytically true serves as a way of blocking a potential escape from this conclusion. One cannot, Maudlin claims, reject RC as a means of avoiding non-locality. Our focus here is restricted to the status of RC, not the thorny issue of quantum non-locality.

Below, we will argue that RC is not analytic. However, there is a deeper worry that we suspect motivates Maudlin's claim:\ that QBism and views like it end up doing away with physical reality altogether. QBism is one of a family of broadly epistemic approaches which regard the quantum state as a reflection of an agent's knowledge, information, or beliefs.\footnote{It should be noted that the broad reading of ``epistemic'' taken here avoids the factive connotation usually associated with knowledge \citep[e.g.,][Sect.\ 2.3]{TimpsonQBism}. Alternatively (and more precisely), one could characterize QBism and related views as \emph{doxastic} approaches to QM (e.g., \citealt[Sect.\ 7.2]{BogeQMBook}; \citealt{deBrotaSIC}).} Other examples include Healey's pragmatist interpretation \citep{HealeyQuantumRevolution}, Friederich's therapeutic view \citep{FriederichTherapeutic}, or Boge's neo-Kantian approach \citep{BogeQMBook}.

These approaches reject RC as part of a more general rejection of the common representational role afforded to QM, a theory which these approaches understand, in the first instance, as a guide for agents navigating the world. The worry is this:\ because of the many known restrictions to supplementing further variables to it \citep[e.g.,][]{BellEPR,PBR2012,BruknerNoGo,FrauchigerRenner}, the abandonment of the representational role of QM seems to imply a form of solipsism or idealism \citep{NorsenQuantumSolipsism, BrownReality}. If RC is taken to be a minimal realist commitment, then its rejection may imply the abandonment of physical reality altogether. We shall argue below that this alternative understanding of RC's status is mistaken as well.

\section{Maudlin's Argument}
\label{Arg}

In defense of his claim that RC is analytic, Maudlin offers the following:

\begin{quote}
...suppose, as the criterion demands, that I can \emph{without in any way disturbing a system} predict with certainty the value of a physical quantity (for example, predict with certainty how the system will react in some experiment). Then, first, there must be some physical fact about the system that determines it will act that way.... Second, if the means of determining this did not disturb the system, then the relevant element of reality obtained \emph{even before} the determination was made, and indeed obtained \emph{independently of the determination being made}. \citep[p.7, original emphasis]{MaudlinBell}
\end{quote}

Maudlin's argument supposes one is in a position to predict with certainty the outcome of some measurement on a system without disturbing the system. That is, there is some physical quantity associated with observable $A$ and one is certain that the outcome of a measurement $O_i$ will yield a particular value of $a$ of $A$: $prob(O_a)=1$.\footnote{A distinction may be drawn between the outcome of a measurement and a physical quantity taking a particular value. As stated, RC concerns only the latter, which may render it inapplicable on certain interpretations of QM \citep[see][p.101]{HealeyQuantumRevolution}.} Maudlin's argument centers on two key claims about such a situation:
 
\begin{enumerate}
    \item There is some physical fact about the system that determines it will behave so as to yield $O_a$.
    \item The physical fact in question is independent of the non-disturbing procedure by which we come to know $prob(O_a)=1$.
\end{enumerate}

This physical fact---the determinant of $O_a$---is the element of reality corresponding to the value $a$ of the physical quantity $A$ that appears in RC. One needn't identify the physical fact with the system's possession of the value $a$ of $A$. So long as there is some physical fact about the system that determines $O_a$ will occur if one measures $A$, one has found a fact corresponding to $A$. On some interpretations, the physical fact in question is the system possessing value $a$ of $A$, but other interpretations will deny this for some observables. For instance, the Bohmian may point to a spatial configuration of particles as the determinant of a measurement concerning an ostensibly non-spatial observable. The intuition here is that without some such prior (or independent) physical fact, we wouldn't be in a position to be \emph{certain} of the measurement outcome.

\section{Objections}
\label{Obj}

\subsection{Subjectivism}

Claim (1) is motivated by the certainty of the outcome $O_a$. In standard QM, the probability of a measurement outcome is given by the quantum state assigned to the system and the Born rule. Here we suppose that the standard formalism of QM delivers $prob(O_a) = 1$---i.e., certainty with respect to finding the system to be $a$. So far so good. But why suppose that this probability claim requires an underlying physical fact that ``determines'' $O_a$?

Indeed, there is long tradition of denying the inference from certainty to reality. The subjectivist (or personalist) school maintains that probability statements correspond to agents' subjective degrees of belief. Thus, $prob(\phi)=1$ means only that one is \emph{certain} of $\phi$, which need not imply that $\phi$ is the case. In the context of QM, subjectivism has been explicitly advocated by QBists \citep{CavesSubjectiveProbability, FuchsMerminSchackIntroQBism}. For the QBist, when QM delivers a probability of 1, this indicates that an agent's beliefs---reflected in the quantum state she ascribes---suggest that she should be certain to experience the outcome in question. But, of course, this doesn't imply that there is some physical fact about the system that ensures the outcome; probability 1 statements are a reflection of agent's degrees of belief just like any other probability statement.

In defense of (1), Maudlin claims that ``if a system is certain to do something physical, then \emph{something} in its physical state entails that it will do it'' \citep[p.7, original emphasis]{MaudlinBell}. However, this begs the question against the subjectivist, who takes ``it is certain that the system will $\phi$'' to mean ``some agent is certain that the system will $\phi$,'' which needn't imply anything about the physical state of the system. %In fact, for the QBist, quantum probabilities concern only our own future experiences, not measurement outcomes understood as objective physical events. Thus, it is also question begging to interpret $prob(O_i)=1$ as certainty of the occurrence of some physical event.

Now, Maudlin may reply that RC's phrase ``we can predict with certainty'' implies not just that one \emph{can} be certain, but that one would be \emph{correct} in doing so. If this is the case, then perhaps we do need some physical state to provide the basis for one's prediction. But, this move is also question begging and, at any rate, not particularly helpful. First, on the classic subjective Bayesian approach which inspires QBism, there are only coherence constraints---i.e., conformity to the probability axioms and Bayesian updating. Thus, it doesn't make sense to speak of a single-case probability as ``correct'' or ``incorrect.'' Second, even if we grant that one is ``correct'' or ``justified'' in being certain that $\phi$ will occur in the future (or would occur if the measurement were performed), this certainty needn't require a physical determinant be present in the system now (or in the actual world). One could be justifiably certain about a future event in an indeterministic universe, in which case there would be no prior determinant of the event \citep[cf.,][pp.38--39]{LewisBell}. Suppose, for instance, that there is an oracle who has direct access to a future measurement event, the outcome of which is the result of a random, indeterministic process. One could be justifiably certain to find $O_a$ without the presence of a feature of the system now that determines the outcome of this future measurement.\footnote{Note that the subjectivist needn't deny the existence of all physical properties of a system to reject RC, but only those that are both (a) attributed on the basis of a probability one statement from QM and (b) determine the outcome of a future measurement. They may, for instance, attribute physical properties to a system associated with its preparation, but such properties will meet neither condition. Preparation procedures may be described without appeal to QM and the laws are indeterministic with respect to preparations and measurement outcomes.}

\subsection{Objectivism}

The objectivist holds that probabilities are genuine features of reality. Even on this understanding, one may wish to resist the move from $prob(O_i)=1$ to $O_i$ being determined to occur. First, indeterministic laws may assign a future event a probability of 1. One may wish to deny that the event in question was \emph{determined} to occur. Second, there are cases which suggest that probability 0 events are not impossible, or determined not to occur. An objectivist persuaded by such cases will deny that probability 1 events must occur, or are determined to occur.\footnote{We assume that if an event is \emph{determined} to occur by some actual event, then it \emph{must} occur. Of course, different laws of nature might allow for a determined event not to occur, so the necessity involved is physical or nomological.}

There are several well-known cases. Consider, for instance, a dart
with a perfectly defined tip thrown at a dartboard. On the standard probability calculus, the probability that it hits a particular point is 0, but it is possible that it does so. Or, consider a fair coin flipped an infinite number of times. The probability of any given sequence---e.g., all heads---is 0, but some such sequence must occur. Some cases can be dealt with by allowing for a non-standard probability measure that assigns infinitesimal probabilities to each event in an infinite outcome space. But, \citet{WilliamsonHeads} argues that this strategy fails in the infinite coin flip case.

%Infinite lotteries present a well-known challenge to any account of probability:\ there seems to be no way for there to be an infinite number of equally-probably tickets whose probabilities sum to 1. There are many potential responses to this challenge, but one compelling option is to deny countable additivity (e.g., \citet{HowsonUrbach}). That is, one denies $\sum_{i=1}^{\infty} prob(W_i) = 1$ for the infinite lottery in question. Taking this option, however, allows for cases where each individual ticket has $prob(W_i)=0$, but the probability of there being some ticket that wins is 1. Hence, there is some infinite collection of 0 probability events which collectively have a probability of 1.

Granting that such cases exist leads to inconsistency if one assumes that $prob(\phi)=1$ ensures that $\phi$ occurs. In particular, this supposition could lead to a situation in which $\forall\phi\neg$($\phi$ will occur) and yet $\exists\phi$($\phi$ will occur). As a result, some will deny the general principle that $prob(\phi)=0$ ensures that $\phi$ will not occur and, correspondingly, that $prob(\phi)=1$ ensures that $\phi$ will occur. This applies just as much to the objectivist about probabilities as to the subjectivist.

An objectivist account of quantum probabilities that squares with these general considerations has been developed, e.g., by \citet{StairsProb1}.\footnote{Thanks to an anonymous referee for bringing this work to our attention.} Stairs advocates a minimal objectivism according to which objective quantum probabilities are grounded in facts about the actual world. However, he rejects the inference from a measurement outcome with objective probability 1 to a property of the target system:\ ``[W]hen Alice makes her probability-one claim about Bob's qubit, she does not need to infer pre-existing properties nor attribute counterfactuals. On the contrary, if she wants to square her objectivism with causal locality, those are exactly the things she should not do'' \citep[p.165]{StairsProb1}. On Stairs' view, quantum theory provides us with objective probabilities in virtue of capturing global statistical patterns, not local intrinsic properties.

\section{Whither Physical Reality?}
Given the difficulties that beset Maudlin's argument, we contend that RC is not analytically true. As a consequence, views that deny RC (e.g., QBism) cannot be dismissed as incoherent on this basis.

However, RC is a modest representational thread linking QM and reality, and hence, one may view its denial as abandonment of the physical world. This suggests a revision of Maudlin's analyticity claim:\ one cannot deny RC without abandoning physical reality altogether.

Indeed, this worry may be thought to be underscored by a number of ``no go'' results that appear to push quantum epistemicists toward ever more subjective positions. Bell's theorem \citep{BellEPR} and more recent results (e.g., \citet{PBR2012,BruknerNoGo,FrauchigerRenner, HealeyObjectivity, LeegwaterGHZ}) are taken by some to require that subjectivism cannot be limited to quantum states, but infects other facets of the epistemicist picture, such as Hamiltonians and measurement results. Thus, one may wonder what is left to form the basis of physical reality. If there are no quantum states, probabilities, or even measurement results in external reality, perhaps nothing remains but the contents of an agent's mind.

However, even if one adopts a thoroughgoing subjectivism about QM, solipsism or idealism need not result. It is perfectly consistent to maintain both that (a) there is a physical reality and (b) that QM doesn't (directly) represent it. Certain epistemic approaches reject RC as a link between theory and reality, but they allow for a more indirect connection between the two. At the very minimum, they may maintain that reality is such that QM is a good guide for the agents who find themselves in it. Such a view may be unsatifisying for one who shares Maudlin's realist inclinations, but it is far from incoherent. In fact, QBism aims to go further than this. For instance, QBists claim that the Born rule ``correlates with something that one might want to call `real'" \citep[p.119]{FuchsParticipatory}. Such claims of objectivity in QBism certainly require some unpacking, but they provide a glimmer of a more indirect route to reality. 

We conclude that opponents of QBism and other epistemic approaches to QM must do more than merely dismiss these views as incoherent or inconsistent with the existence of physical reality.

\paragraph*{Acknowledgements} F. J. Boge was employed with the research unit \emph{The Epistemology of the Large Hadron Collider} (German Research Foundation (DFG); grant FOR 2063) while working on this paper, and hence acknowledges the funding. We otherwise thank Richard Healey and John DeBrota for helpful comments on an earlier version of this paper, as well as Chris Fuchs for discussions on the subject at the Stellenbosch Institute for Advanced Study. 
\bibliographystyle{apalike}
\bibliography{references}

\end{document}